\documentclass[letterpaper, 10 pt, conference]{ieeeconf}  

\IEEEoverridecommandlockouts                              

\usepackage[colorlinks,urlcolor=blue,linkcolor=blue,citecolor=blue]{hyperref}
\usepackage{color,array}
\usepackage{graphicx}

\usepackage[utf8]{inputenc}
\usepackage[T1]{fontenc}

\usepackage{amsmath}
\usepackage{amsfonts}
\usepackage{amssymb}
\hypersetup{colorlinks=true, linkcolor=blue, filecolor=magenta, urlcolor=cyan,}
\usepackage{mathrsfs}
\usepackage{bbold}

\usepackage[export]{adjustbox}
\usepackage{todonotes}

\usepackage{algorithm}
\usepackage{algpseudocode}

\usepackage{subcaption}

\newlength{\whilewidth}
\algdef{SE}[parWHILE]{parWhile}{EndparWhile}[1]
{\parbox[t]{\dimexpr\linewidth-\algmargin}{%
		\hangindent\whilewidth\strut\algorithmicwhile\ #1\ \algorithmicdo\strut}}{\algorithmicend\ \algorithmicwhile}

\title{\LARGE \bf
Statistical Verification of Traffic Systems with Expected Differential Privacy
}

\author{Mark Yen$^{1}$, Geir E. Dullerud$^{2}$, and Yu Wang$^{1}$
\thanks{$^{1}$Mark Yen and Yu Wang are with the Department of Mechanical \& Aerospace Engineering, University of Florida, Gainesville, FL, USA. Mark Yen is partly supported by the DoD SMART Scholarship. Email: {\tt\small \{markyen, yuwang1\}@ufl.edu}}
\thanks{$^{2}$Geir E. Dullerud is with the Department of Mechanical Science \& Engineering, University of Illinois at Urbana-Champaign, Urbana, IL, USA. Email: {\tt\small dullerud@illinois.edu}}
}

\newtheorem{definition}{Definition}
\newtheorem{assumption}{Assumption}
\newtheorem{theorem}{Theorem}

\newcommand{\nat}{\mathbb{N}}
\newcommand{\real}{\mathbb{R}}
\newcommand{\nnreal}{\real_{\geq 0}}
\newcommand{\algmargin}{\the\ALG@thistlm}

\algnewcommand{\LineComment}[1]{\Statex \(\triangleright\) #1}
\algnewcommand{\parState}[1]{\State%
	\parbox[t]{\dimexpr\linewidth-\algmargin}{\strut #1\strut}}
\algnewcommand{\parRequire}[1]{\Require%
	\parbox[t]{\dimexpr\linewidth-\algmargin}{\strut #1\strut}}

\begin{document}

\maketitle
\thispagestyle{empty}
\pagestyle{empty}

\begin{abstract}
Traffic systems are multi-agent cyber-physical systems whose performance is closely related to human welfare. They work in open environments and are subject to uncertainties from various sources, making their performance hard to verify by traditional model-based approaches. Alternatively, statistical model checking (SMC) can verify their performance by sequentially drawing sample data until the correctness of a performance specification can be inferred with desired statistical accuracy. This work aims to verify traffic systems with privacy, motivated by the fact that the data used may include personal information (e.g., daily itinerary) and get leaked unintendedly by observing the execution of the SMC algorithm. To formally capture data privacy in SMC, we introduce the concept of expected differential privacy (EDP), which constrains how much the algorithm execution can change in the expectation sense when data change. Accordingly, we introduce an exponential randomization mechanism for the SMC algorithm to achieve the EDP. Our case study on traffic intersections by Vissim simulation shows the high accuracy of SMC in traffic model verification without significantly sacrificing computing efficiency. The case study also shows EDP successfully bounding the algorithm outputs to guarantee privacy.
\end{abstract}


\section{Introduction}
\label{sec:intro}

Traffic systems are important multi-agent cyber-physical systems closely related to human welfare. They are usually influenced by uncertainties and even adversarial attacks from various sources. Therefore, it is vital to verify traffic system performance to ensure safety and efficiency. Previously, the main verification approach for traffic systems was based on model checking. In recent literature, various model checking methods have been developed using abstraction~\cite{roohi2017hare}, linear temporal logic~\cite{coogan2015efficient, alur2015principles}, and symmetric~\cite{sibai2019using}, to name a few.

With an increasing abundance of data, there is a growing interest in data-driven verification methods, particularly statistical model checking (SMC). SMC can verify general specifications expressible by temporal logic for a wide range of systems. SMC works by just sampling the system, which allows it to handle black-box systems that are either discrete, continuous, or hybrid~\cite{agha2018survey}. It also allows SMC to be more scalable than other model-based verification methods that suffer from the curse of dimensionality~\cite{younes2006numerical}.

However, using SMC to verify traffic systems raises privacy concerns because intruders can observe the SMC outputs and the number of samples used to infer the values of the system samples~\cite{dwork2006differential}.
For instance, consider an algorithm that seeks to measure and control the speed and direction of vehicles driving through a traffic intersection. Using SMC to verify the algorithm (e.g., to assure performance in rush hours) may compromise the drivers' privacy if intruders could infer the vehicle trajectories by observing SMC outputs and termination times.

A promising method to protect data privacy in SMC is to use differential privacy. Differential privacy is achieved when a random observation of an algorithm does not change significantly after small changes are made to the sample values by a carefully designed randomization mechanism~\cite{dwork2006differential}. If a bound exists on the change, it will guarantee how difficult it is to infer the sample values from the algorithm's observations.
Previous work focused on an algorithm's output sensitivity to changes in the input data~\cite{dwork2013algorithmic}. However, SMC is a sequential algorithm where the number of samples collected (i.e., sample termination time) depends on the termination condition as well as the input data. Thus, we must consider both sample termination time and output sensitivities when using SMC. Previous work on differential privacy for sequential algorithms~\cite{ghassemi2016differentially,jain2011differentially,tsitsiklis2018private} usually avoids this obstacle by assuming that the sample termination time is not observable by the intruder.

Since standard differential privacy is challenging (if not impossible) to achieve for sequential algorithms like SMC, a new definition called \emph{expected differential privacy} (EDP) is introduced \cite{wangDifferentiallyPrivateAlgorithms2022a}.
EDP utilizes an assumption that is not previously considered in standard differential privacy. More specifically, EDP relies on the idea that SMC draws independently and identically distributed (i.i.d.) samples from a probabilistic model, and the distribution of the sample sequences will converge to the underlying distribution of the model. Achieving EDP means that for any \textit{distribution} of sequences of samples, changing the value of the same singular sample in each sequence will only slightly change the \textit{average} of the algorithm output.
This is different from standard differential privacy, which states that for any values of a sequence of samples, changing the value of one sample will only slightly change the algorithm output.

\begin{figure}[t]
    \vspace{5pt}
    \centering
    \includegraphics[width=0.9\linewidth]{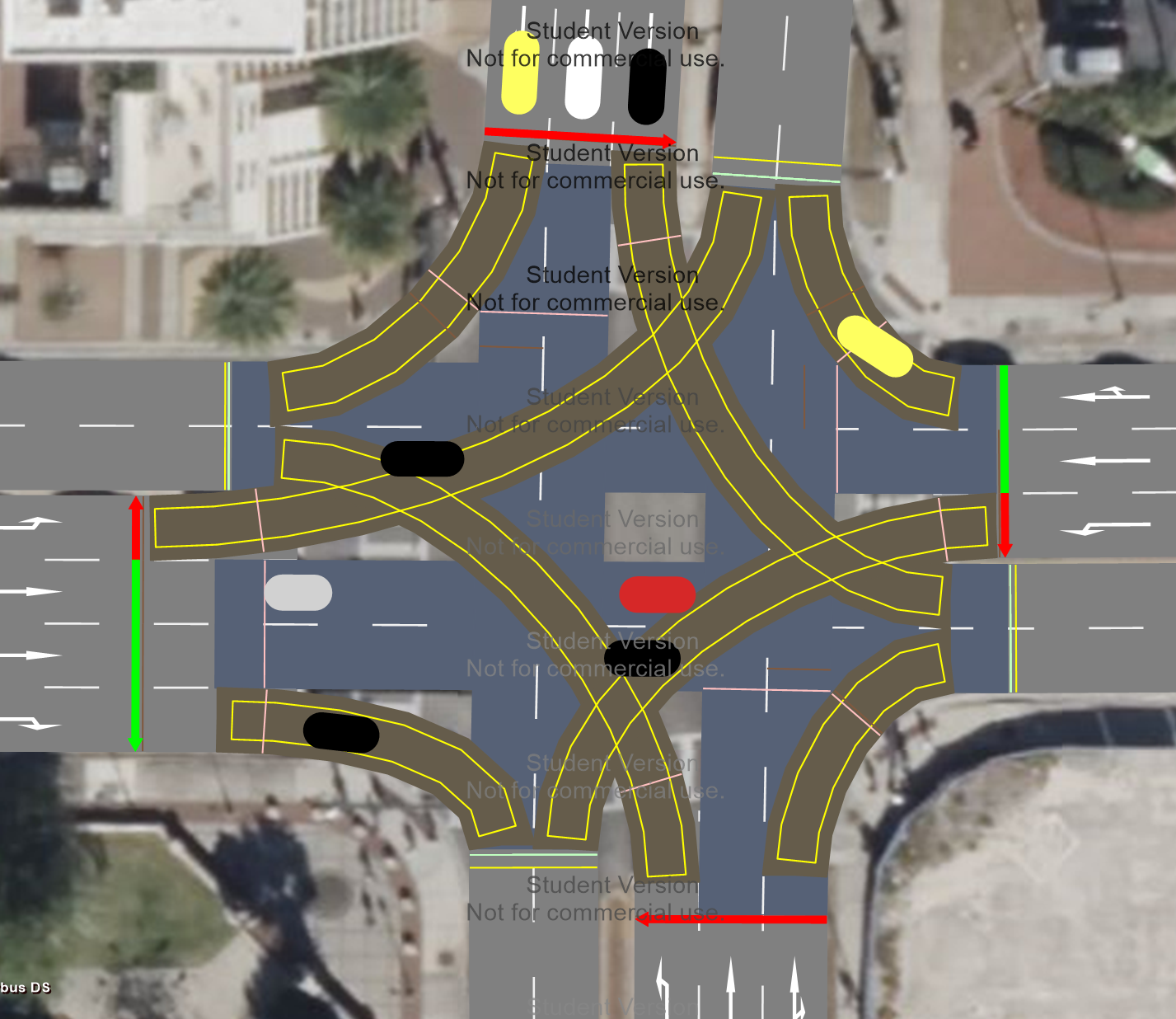}
    \caption{\footnotesize Screenshot of the intersection at West University Avenue and 13th Street at Gainesville, FL, USA modeled by PTV Vissim simulator. It is a multi-agent system where each vehicle is an agent that follows, among other factors, a probability distribution of driving decisions, a Gaussian distribution of driving speeds, and the inherent driving dynamics. Vehicle speeds and driving decisions were recorded over a simulation time horizon of $T \in \real$ seconds for a total of $10^5$ samples.}
    \label{fig:vissim scene}
    \vspace{-13pt}
\end{figure}

The main contribution of this work is to implement the SMC with EDP algorithm from \cite{wangDifferentiallyPrivateAlgorithms2022a} to analyze a model of an intersection at the University of Florida simulated by PTV Vissim, as shown in Fig. \ref{fig:vissim scene}. The results show that SMC can accurately verify the percentage of vehicles that fall within a range of acceptable speeds while drawing a moderate number of samples from the system. The upper speed limit ensures that vehicles do not endanger pedestrians, while the lower speed limit encourages efficient traffic flow. We also show that the EDP effectively bounds the difference between two distributions of randomized SMC outputs, thus guaranteeing privacy.

The rest of the paper is organized as follows.
Section~\ref{sec:smc} provides some preliminaries on SMC.
Section~\ref{sec:edp} introduces EDP and contrasts it with standard differential privacy.
Section~\ref{sec:smc edp} introduces SMC with EDP and presents a pseudo code of the algorithm, based on \cite{wangDifferentiallyPrivateAlgorithms2022a}.
Section~\ref{sec:case} uses SMC with EDP on a PTV Vissim traffic model as a case study and discusses the findings.
Finally, Section~\ref{sec:conc} concludes this work.

\paragraph*{Notations} The set of natural, real, and non-negative real numbers are denoted by $\nat$, $\real$ and $\nnreal$, respectively. For $n \in \nat$, let $[n] = \{1, \ldots, n\}$. The binomial distribution has a probability mass function in the following form
$$f(K; N, p) = \frac{N!}{K! (N-K)!} p^K (1-p)^{N-K},$$ 
and is denoted by $\mathrm{Binom} (N, p)$, where $K, N \in \nat$ and $p$ is a real value within the interval $[0,1]$. The exponential distribution has a probability density function of the form 
$$
f(x) = \begin{cases}
\varepsilon e^{- \varepsilon x}, & x \geq 0 \\
0, & x < 0 \label{eq:L}
\end{cases}
$$
and is denoted by $\mathrm{Exp} (\varepsilon)$, where $\varepsilon \in \nnreal$ and $x \in \real$.




\section{Statistical model checking} \label{sec:smc}

This section provides preliminaries on statistical model checking. Consider a random signal $\sigma$ from an arbitrary model $\mathcal{S}$. The goal of SMC is to check the probability of $\sigma$ satisfying a specification $\varphi$ being greater than some given threshold $p \in [0,1]$,
\begin{equation} \label{eq: SMC}
	\mathbb{P}_{\sigma \sim \mathcal{S}} ( \sigma \models \varphi ) = p_\varphi > p,
\end{equation}
where $\models$ means ``to satisfy'' and $p_\varphi$ is the satisfaction probability of $\varphi$ on $\mathcal{S}$. 

Formally, $\varphi$ is expressed by signal temporal logic (STL)~\cite{maler2004monitoring}, which is an extension of linear temporal logic to systems with real-valued time and states. STL specifications are defined recursively by:
\[
\varphi :=
f (\sigma) \geq 0
\mid \neg \varphi
\mid \varphi \land \varphi
\mid \varphi \ \mathcal{U}_{[t_1,t_2]} \ \varphi,
\]
where $\sigma: \nnreal \to \real^n$ is a real-valued signal
, $f: \real^n \to \real$ is a real-valued function of the signal, and $[t_1, t_2]$ is a time interval with $t_2 > t_1 \geq 0$ and $t_1$ and $t_2$ being rational-valued or infinity. 
For any STL specification $\varphi$, we can define whether a signal $\sigma$ satisfies it or not, written as $\sigma \models \varphi$ or $\sigma \not\models \varphi$.
Specifically, $\sigma \models f (\sigma) \geq 0$ if and only if $f(\sigma(0)) \geq 0$. This rule means that $\sigma$ satisfies $f (\sigma) \geq 0$ if and only if it holds at time $0$.
In addition, $ \sigma \models \varphi \ \mathcal{U}_{[t_1,t_2]} \ \psi$ if and only if $\exists t \in [t_1, t_2] \text{ such that } \sigma^{(t)} \models \varphi$ and $\forall t' < t, \sigma^{(t')} \models \varphi$ where $\sigma^{(t)}$ denotes the $t$-shift of $\sigma$, defined by $\sigma^{(t)}(t') = \sigma(t + t')$ for any $t' \in \nnreal$.
This rule defined the ``until'' temporal operator, meaning that $\sigma$ satisfies $\varphi \ \mathcal{U}_{[t_1,t_2]} \ \psi$ if and only if $\sigma$ satisfies $\varphi$ at all time instants before $t$ until it satisfies $\psi$ exactly at time $t$ for some $t \in [t_1,t_2]$.
Finally, we can define other temporal operators such as ``finally'' (or ``eventually'') and ``always'', written as $\Diamond$ and $\square$. Specifically,
$\Diamond_{[t_1,t_2]} \varphi = \mathtt{True} \ \mathcal{U}_{[t_1,t_2]} \ \varphi$
means the property $\varphi$ finally holds; and 
$\square_{[t_1,t_2]} \varphi = \neg (\Diamond_{[t_1,t_2]} \neg\varphi)$ 
means the property $\varphi$ always holds.






SMC handles \eqref{eq: SMC} as a hypothesis testing problem
\begin{align} \label{eq: hypothesis testing}
H_\textrm{null}: p_\varphi > p, \quad
H_\textrm{alt}: p_\varphi \leq p, 
\end{align}
where $H_\textrm{null}$ and $H_\textrm{alt}$ are the null and alternative hypotheses respectively. Then, it draws sample signals $\sigma_1, \sigma_2, \ldots$ from $\mathcal{S}$ to verify whether $H_\textrm{null}$ or $H_\textrm{alt}$ holds. Correlated sampling has been shown to improve SMC efficiency, but it requires some knowledge of the system dynamics~\cite{wang2018statisticala,wang2019statistical}. A traffic intersection is a highly-complex system that depends on many factors. Therefore, this work focuses on independent sampling, which can handle general black-box systems~\cite{agha2018survey, legay2010statistical}.


SMC examines the correctness of $\varphi$ on each sample signal $\sigma_i$ as the following Boolean
\begin{equation} \label{eq:varphi}
\varphi(\sigma_i) = \begin{cases}
1, & \sigma_i \models \phi, \\
0, &\text{ otherwise.}
\end{cases}
\end{equation}
Since the sample signals are independent, the sum 
\begin{equation} \label{eq:K}
	K = \sum_{i \in [N]} \varphi (\sigma_{i})
\end{equation}
is a random variable following the binomial distribution $\mathrm{Binom} (N, p_\varphi)$, so the following approximation can be made
\begin{equation}
    \label{eq: p_phi approx}
    p_\varphi \approx \frac{K}{N}.
\end{equation}

SMC aims to find a statistical assertion 
$\mathscr{A} \big( \sigma_1, \ldots, \sigma_N \big) \allowbreak \to \{H_\textrm{null}, H_\textrm{alt}\}$
that claims either $H_\textrm{null}$ or $H_\textrm{alt}$ holds based on the observed samples. Since $K$ is the sufficient statistics, the statistical assertion can be written as 
$$\mathscr{A} \big(K, N) \to \{H_\textrm{null}, H_\textrm{alt}\}.$$ 
Due to the randomness of $\sigma_1, \ldots, \sigma_N$, the value of $\mathscr{A}$ does not always agree with the truth value of $\mathbb{P}_{\sigma \sim \mathcal{S}} (\sigma \models \varphi) < p$. Thus, to capture these probabilistic errors, the false positive/false negative (FP/FN) ratios can be defined as
\begin{align} 
& \alpha_\textrm{FP} = \mathbb{P}_{\sigma_1, \ldots, \sigma_N \sim \mathcal{S}} \big( \mathscr{A} = H_\textrm{alt} \mid \mathbb{P}_{\sigma \sim \mathcal{S}} (\sigma \models \varphi) > p \big), \label{eq:fp_def}
\\ & \alpha_\textrm{FN} = \mathbb{P}_{\sigma_1, \ldots, \sigma_N \sim \mathcal{S}} \big( \mathscr{A} = H_\textrm{null} \mid \mathbb{P}_{\sigma \sim \mathcal{S}} (\sigma \models \varphi) \leq p \big). \label{eq:fn_def}
\end{align}
The FP ratio is the error probability of rejecting $H_\textrm{alt}$ when~\eqref{eq: SMC} holds and the FN ratio is the error probability of accepting $H_\textrm{null}$ when~\eqref{eq: SMC} does not hold.

Assertion $\mathscr{A}$ becomes more accurate, in terms of decreasing $\alpha_\textrm{FP}$ and $\alpha_\textrm{FN}$, when the number of samples $N$ increases. For any $N$, quantitative bounds of $\alpha_\textrm{FP}$ and $\alpha_\textrm{FN}$ can be derived by using either the confidence interval method~\cite{zarei2020statistical}, which only assumes $p_\varphi \neq p$, or the sequential probability ratio test method~\cite{sen2004statistical}, which requires a stronger assumption but is more efficient. This work focuses on the latter method since the following indifference parameter assumption holds in most applications, such as in~\cite{roohi2017statistical}.

\begin{assumption} \label{as:indifference}
There exists a known indifference parameter $\delta > 0$ such that $|p_\varphi - p| > \delta$ in~\eqref{eq: SMC}.
\end{assumption}

With Assumption~\ref{as:indifference}, we can consider two extreme cases in \eqref{eq: hypothesis testing}, according to~\cite{sen2004statistical},
\begin{align} \label{eq:problem 1 sht}
H_\textrm{null}: p_\varphi = p + \delta, \quad H_\textrm{alt}: p_\varphi = p - \delta.
\end{align}
To distinguish between $H_\textrm{null}$ or $H_\textrm{alt}$, we consider the likelihood ratio
\begin{equation} \label{eq:probability_ratio}
\lambda(K,N) = \frac{ (p+\delta)^{K} (1-p-\delta)^{N-K} }{ (p-\delta)^{K} (1-p+\delta)^{N-K} }.
\end{equation}
SMC aims to ensure $\alpha_\textrm{FP}, \alpha_\textrm{FN} \leq \alpha$ for a desired significance level $\alpha > 0$.
Using sequential probability ratio tests~\cite{casella2002statistical}, SMC draws system samples until either condition on the right-hand-side of the following equation is satisfied before making a statistical assertion:
\begin{equation} \label{eq:sprt_assertion}
\mathscr{A} \big(K, N) = \begin{cases}
	H_\textrm{null}, &\text{ if } \lambda(K,N) \geq \frac{1 - \alpha}{\alpha}, \\
	H_\textrm{alt}, &\text{ if } \lambda(K,N) \leq \frac{\alpha}{1 - \alpha}.
\end{cases}
\end{equation}
As $N \to \infty$, by the binomial distribution, we have $\lambda(K,N) \to 0$ if $H_\textrm{null}$ holds or $\lambda(K,N) \to \infty$ if $H_\textrm{alt}$ holds, so \eqref{eq:sprt_assertion} stops with probability $1$.
\section{Expected differential privacy for sequential algorithms} \label{sec:edp}

This section formally introduces expected differential privacy for sequential algorithms, based on \cite{wangDifferentiallyPrivateAlgorithms2022a}.
Consider a sequential algorithm $\mathscr{B}$ (e.g., SMC). It takes sample signals $\sigma_{1:\infty} = (\sigma_1, \sigma_2, \ldots)$ from a probabilistic model $\mathcal{S}$ until its termination condition is satisfied. We denote the termination time (i.e., the number of samples used) by $\tau_\mathscr{B} \in \nat$ and the final algorithm output by $o_\mathscr{B}$. The termination time and output are functions of $\sigma_{1:\infty}$ with $\tau_\mathscr{B}  (\sigma_{1:\infty})$ and $o_\mathscr{B}  (\sigma_{1:\infty})$, respectively.

For any two input sequences $\sigma_{1:\infty}$ and $\sigma_{1:\infty}'$ differing in only one sample $\sigma_i$, standard differential privacy requires $\mathscr{B}$ to satisfy the following
\begin{align} \label{eq:standard differential privacy}
&\mathbb{P}_{\mathscr{B}} \Big( \big( \tau_{\mathscr{B}} (\sigma_{1:\infty}), o_{\mathscr{B}} (\sigma_{1:\infty}) \big) \in \mathcal{O} \Big) \notag
\\ &\leq e^{\varepsilon} \mathbb{P}_{\mathscr{B}} \Big( \big( \tau_{\mathscr{B}} (\sigma_{1:\infty}'), o_\mathscr{B} (\sigma_{1:\infty}') \big) \in \mathcal{O} \Big)
\end{align}
for any $\mathcal{O} \subseteq \mathbb{N} \times \{H_\textrm{null}, H_\textrm{alt}\}$. However, it has been shown in \cite{wangDifferentiallyPrivateAlgorithms2022a} that the difference in the termination times $\tau_{\mathscr{B}} (\sigma_{1:\infty})$ and $\tau_{\mathscr{B}} (\sigma_{1:\infty}')$ can approach $\infty$. Thus, \eqref{eq:standard differential privacy} cannot be satisfied with a finite value of $\varepsilon$.
Instead, the following definition for expected differential privacy can be used:

\begin{definition}[\cite{wangDifferentiallyPrivateAlgorithms2022a}] \label{def:expected differential privacy}
A randomized sequential algorithm $\mathscr{B}$ is $\varepsilon$-expectedly differentially private if
\begin{align} \label{eq:EDP}
&\mathbb{P}_{\mathscr{B}} \Big( \mathbb{E}_{\sigma_{-n}} \big( \tau_{\mathscr{B}} (\sigma_{1:\infty}), o_{\mathscr{B}} (\sigma_{1:\infty}) \big) \in \mathcal{O} \Big) \notag
\\ & \leq e^{\varepsilon} \mathbb{P}_{\mathscr{B}} \Big( \mathbb{E}_{\sigma'_{-n}} \big( \tau_{\mathscr{B}} (\sigma_{1:\infty}'), o_{\mathscr{B}} (\sigma_{1:\infty}') \big) \in \mathcal{O} \Big),
\end{align}
for any $n \in \nat$, any set $\mathcal{O}$, and any pair of signals $\sigma_n$ and $\sigma'_n$, where $\sigma_{-n}$ and $\sigma'_{-n}$ are random sequences that exclude the $n \text{-th}$ pair of signals and contain i.i.d. entries following the probabilistic model $\mathcal{S}$.
\end{definition}

This new notion of differential privacy relies on the following assumption that has not been considered in~\eqref{eq:standard differential privacy}: SMC draws i.i.d. samples from an underlying probability distribution of system $\mathcal{S}$.
This assumption is stronger than the arbitrarily-valued static databases assumed in~\cite{dwork2008differential}. Thus, it allows a relaxed notion of privacy to be defined based on bounding the change of the {\em average} of the algorithm's output over the {\em distribution} of $\sigma_{1:\infty}$ with respect to arbitrary changes of an arbitrary {\em entry} $\sigma_n$. In contrast, standard differential privacy is based on bounding the change of the algorithm's output for {\em any} $\sigma_{1:\infty}$ with respect to arbitrary changes of an arbitrary {\em signal} $\sigma_i$.

Definition \ref{def:expected differential privacy} is weaker than the standard differential privacy definition because the condition of~\eqref{eq:standard differential privacy} implies \eqref{eq:EDP} by taking the expected values $\mathbb{E}_{\sigma_{-n}} \mathbb{E}_{\sigma'_{-n}}$ on both sides of~\eqref{eq:standard differential privacy}, i.e.,
\begin{align*}
& \mathbb{E}_{\sigma_{-n}} \mathbb{E}_{\sigma'_{-n}} \mathbb{P}_{\mathscr{B}} \Big( \big( \tau_{\mathscr{B}} (\sigma_{1:\infty}), o_{\mathscr{B}} (\sigma_{1:\infty}) \big) \in \mathcal{O} \Big)
\\ & \leq e^{\varepsilon} \mathbb{E}_{\sigma_{-n}} \mathbb{E}_{\sigma'_{-n}} \mathbb{P}_{\mathscr{B}} \Big( \big( \tau_{\mathscr{B}} (\sigma_{1:\infty}'), o_{\mathscr{B}} (\sigma_{1:\infty}') \big) \in \mathcal{O} \Big),
\end{align*}
for any $\mathcal{O} \subseteq \mathbb{N}$, $\sigma_n$, and $\sigma_n'$. This implies that
\begin{align*}
& \mathbb{E}_{\sigma_{-n}} \mathbb{P}_{\mathscr{B}} \Big( \big( \tau_{\mathscr{B}} (\sigma_{1:\infty}), o_{\mathscr{B}} (\sigma_{1:\infty}) \big) \in \mathcal{O} \Big)
\\ & \leq e^{\varepsilon} \mathbb{E}_{\sigma'_{-n}} \mathbb{P}_{\mathscr{B}} \Big( \big( \tau_{\mathscr{B}} (\sigma_{1:\infty}'), o_{\mathscr{B}} (\sigma_{1:\infty}') \big) \in \mathcal{O} \Big).
\end{align*}
Note that 
\begin{align*}
& \mathbb{E}_{\sigma_{-n}} \mathbb{P}_{\mathscr{B}} \Big( \big( \tau_{\mathscr{B}} (\sigma_{1:\infty}), o_{\mathscr{B}} (\sigma_{1:\infty}) \big) \in \mathcal{O} \Big)
\\ & = \mathbb{E}_{\sigma_{-n}} \mathbb{E}_{\mathscr{B}} \ \textbf{I} \Big( \big( \tau_{\mathscr{B}} (\sigma_{1:\infty}), o_{\mathscr{B}} (\sigma_{1:\infty}) \big) \in \mathcal{O} \Big)
\\ & = \mathbb{E}_{\mathscr{B}} \mathbb{E}_{\sigma_{-n}}  \ \textbf{I} \Big( \big( \tau_{\mathscr{B}} (\sigma_{1:\infty}), o_{\mathscr{B}} (\sigma_{1:\infty}) \big) \in \mathcal{O} \Big)
\\ & \leq e^{\varepsilon} \mathbb{E}_{\sigma'_{-n}} \mathbb{P}_{\mathscr{B}} \Big( \big( \tau_{\mathscr{B}} (\sigma_{1:\infty}'), o_{\mathscr{B}} (\sigma_{1:\infty}') \big) \in \mathcal{O} \Big),
\end{align*}
where $\textbf{I}(\cdot)$ is the indicator function and $\mathbb{E}_{\sigma_{-n}}$ and $\mathbb{E}_{\mathscr{B}}$ commute by Fubini's Theorem. Thus, $\mathbb{E}_{\sigma_{-n}}$ and $\mathbb{P}_{\mathscr{B}}$ (or $\mathbb{E}_{\sigma'_{-n}}$ and $\mathbb{P}_{\mathscr{B}}$) can be flipped to arrive back at~\eqref{eq:EDP}.
\section{Statistical model checking with expected differential privacy} \label{sec:smc edp}

This section discusses how EDP is incorporated into SMC \cite{wangDifferentiallyPrivateAlgorithms2022a}.
SMC yields observations on the termination time $\tau_{\mathscr{A}}$ and output $o_{\mathscr{A}}$. Achieving EDP with respect to $o_{\mathscr{A}}$ is trivial, thus the discussion focuses on the termination time $\tau_\mathscr{A}$. It depends on the likelihood ratio $\lambda(N,K)$ in \eqref{eq:probability_ratio}, thus we define the log-likelihood ratio
\begin{equation} \label{eq:llr}
\Lambda_N = \ln \lambda(K,N) = K s_+ - (N - K) s_-, \text{ where}
\end{equation}
$$
s_+ := \ln \frac{ p+\delta }{ p-\delta } > 0 \text{ and} \quad s_- := \ln \frac{ 1-p+\delta }{ 1-p-\delta } > 0
$$
Using \eqref{eq: p_phi approx}, it is seen that \eqref{eq:llr} forms an asymmetric random walk with probabilities and step sizes being $(p_\varphi, s_+)$ and $(1 - p_\varphi, -s_-)$. Thus, \eqref{eq:sprt_assertion} can be seen as a random walk that terminates at the following upper or lower bounds
\begin{align} \label{eq:stopping condition}
    B_+ := \ln \frac{1-\alpha}{\alpha} > 0, \quad B_- := - \ln \frac{1-\alpha}{\alpha} < 0.
\end{align}
Within the bounds, the average step size of the random walk is
\vspace{-5pt}
\begin{align} \label{eq:average step size}
D := & \mathbb{E}_{\sigma_1} \Lambda_1 = p_\varphi s_+ - (1 - p_\varphi) s_- \notag
\\ & = p_\varphi \ln \frac{ p+\delta }{ p-\delta } - (1 - p_\varphi) \ln \frac{ 1-p+\delta }{ 1-p-\delta }.
\end{align}
Using \eqref{eq:average step size}, we show that the average termination time $\tau_{\mathscr{A}} (\sigma_{1:\infty})$ satisfies the stopping time property of random processes~\cite{casella2002statistical}, i.e., 
\begin{align} \label{eq:stopping time property}
\mathbb{E}_{\sigma_{1:\infty}} [\Lambda_{\tau_{\mathscr{A}} (\sigma_{1:\infty})}] 
& = \mathbb{E}_{\sigma_1} \Lambda_{1} \mathbb{E}_{\sigma_{1:\infty}} [ \tau_{\mathscr{A}} (\sigma_{1:\infty}) ] \notag
\\ & = D \mathbb{E}_{\sigma_{1:\infty}} [ \tau_{\mathscr{A}} (\sigma_{1:\infty}) ].
\end{align}

We can use sensitivity analysis to achieve EDP for the SMC algorithm using \eqref{eq:stopping time property}. First, we define the sensitivity of the average termination time $\tau_\mathscr{A} (\sigma_{1:\infty})$ of SMC to changes in an arbitrary signal. Instead of the standard sensitivity \cite{mcsherry2007mechanism}, we use the {\em expected} sensitivity of the termination time $\Delta_{\tau_\mathscr{A}}$ so that the analysis aligns with the definition of EDP:
\begin{equation} \label{eq:expected sensitivity}
\small
\Delta_{\tau_\mathscr{A}} := \max_{n \in \mathbb{N}, \sigma_n, \sigma'_n} \big| \mathbb{E}_{\sigma_{-n}} \tau_\mathscr{A} (\sigma_{1:\infty}) - \mathbb{E}_{\sigma_{-n}'} \tau_\mathscr{A} (\sigma_{1:\infty}') \big|,
\end{equation}
where $\sigma_{-n}$ and $\sigma_{-n}'$ are two random sequences of signals that exclude the $n \text{-th}$ pair of signals and follow the distribution of probabilistic model $\mathcal{S}$.

To calculate \eqref{eq:expected sensitivity}, we rearrange \eqref{eq:stopping time property} and analyze the expected random walk $\mathbb{E}_{\sigma_{1:\infty}} [\Lambda_{\tau_{\mathscr{A}} (\sigma_{1:\infty})}]$.
For $D > 0$, 
consider the random walk $\Lambda_N$ from \eqref{eq:llr} for hitting any two absorbing bounds $A \gg \max\{s_+, s_-\}$ and $-B \ll - \max\{s_+, s_-\}$. The probability of hitting $B$ is $\approx e^{-B}$ so 
the expected termination time becomes
\begin{equation} \label{eq: expected termination time}
\mathbb{E}_{\sigma_{1:\infty}} [ \tau_{\mathscr{A}} (\sigma_{1:\infty}) ] \approx \frac{ A (1 - e^{-B})  - B e^{-B} }{D}.
\end{equation}



Afterwards, using \eqref{eq: expected termination time}, the expected sensitivity of the termination time $\tau_{\mathscr{A}}$ satisfies
\begin{equation} \label{eq:Delta}
\small
\Delta_{\tau_\mathscr{A}} = \frac{s_+ + s_-}{|D|}
= \frac{\ln \frac{ p+\delta }{ p-\delta } + \ln \frac{ 1-p+\delta }{ 1-p-\delta }}{\big| p_\varphi \ln \frac{ p+\delta }{ p-\delta } - (1 - p_\varphi) \ln \frac{ 1-p+\delta }{ 1-p-\delta } \big|}.
\end{equation}
If $\Delta_{\tau_\mathscr{A}}$ is finite, then the exponential mechanism for standard differential privacy can be modified into the following new exponential mechanism to help SMC achieve EDP:
\begin{align} \label{eq:exponential mechanism 2}
& \mathbb{P}_{\mathscr{B}} \big( \mathbb{E}_{\sigma_{1:\infty}} \big[ \tau_\mathscr{B} (\sigma_{1:\infty})\big] = k \big) \notag
\\ &  
\qquad = \frac{e^{- \varepsilon |k - \mathbb{E}_{\sigma_{1:\infty}} [\tau_\mathscr{A} (\sigma_{1:\infty})]| / \Delta_{\tau_\mathscr{A}}}}{\sum_{h \in \nat} e^{- \varepsilon |h - \mathbb{E}_{\sigma_{1:\infty}} [\tau_\mathscr{A} (\sigma_{1:\infty})]| / \Delta_{\tau_\mathscr{A}}}}.
\end{align}
Recall that sequential algorithm $\mathscr{A}$ has a deterministic termination time $\tau_\mathscr{A}$ and sequential algorithm $\mathscr{B}$ has the same input and output spaces as $\mathscr{A}$, but has a random termination time $\tau_{\mathscr{B}}$. Thus, \eqref{eq:exponential mechanism 2} implies $\varepsilon$-expectedly differentially private.

To incorporate \eqref{eq:exponential mechanism 2} into SMC, the randomization was applied to the upper and lower bounds in \eqref{eq:stopping condition} instead of the SMC output. This is because adjusting the termination condition is more straightforward than randomizing the termination output. Thus, the stopping condition \eqref{eq:stopping condition} will be modified into the following
\begin{align} \label{eq:randomized stopping condition}
& B_+ = \ln \frac{1-\alpha}{\alpha} + L \text{ and} \quad B_- = - \ln \frac{1-\alpha}{\alpha} - L, \text{ where }  \notag 
\\ & L \sim \textrm{Exp} \bigg( \frac{\varepsilon}{\ln \frac{ p+\delta }{ p-\delta } + \ln \frac{ 1-p+\delta }{ 1-p-\delta }} \bigg).
\end{align}
Accordingly, we present the SMC with EDP by Algorithm~\ref{alg:2}. The following theorem from \cite{wangDifferentiallyPrivateAlgorithms2022a} guarantees the EDP and statistical accuracy for Algorithm~\ref{alg:2}. 
\begin{theorem}
Algorithm~\ref{alg:2} is $2 \varepsilon$-expectedly differentially private and has a significance level (i.e., an upper bound on the probability that this algorithm returns a wrong answer) less than $\alpha$.
\end{theorem}

\begin{algorithm}[b]
\caption{Statistical model checking of $\mathbb{P}_{\sigma \sim \mathcal{S}} (\sigma \models \varphi) < p$ with expected differential privacy.\label{alg:2}}

\begin{algorithmic}[1]
\Require Probabilistic model $\mathcal{S}$, desired significance level $\alpha$, and indifference parameter $\delta$, privacy level $\varepsilon$.

\State $N \gets 0$, $K \gets 0$.

\State Log-likelihood ratio $\Lambda \gets 0$.

\State $B \gets \ln \frac{1-\alpha}{\alpha}$, $L \sim \textrm{Exp} \Big( \frac{\varepsilon}{\ln \frac{ p+\delta }{ p-\delta } + \ln \frac{ 1-p+\delta }{ 1-p-\delta }} \Big)$

\While{True}

\State Draw a sample signal $\sigma$ from $\mathcal{S}$.

\State $K \gets K + \varphi(\sigma)$, $N \gets N + 1$.

\State $\Lambda \gets \Lambda + \ln \frac{ (p+\delta)^{\varphi(\sigma)} (1-p-\delta)^{1-\varphi(\sigma)} }{ (p-\delta)^{\varphi(\sigma)} (1-p+\delta)^{1-\varphi(\sigma)} }$ .

\If {$\Lambda \geq B + L$} 

\State Return $H_\textrm{null}$ 

\ElsIf {$\Lambda \leq -B - L$} 

\State Return $H_\textrm{alt}$

\Else \ Continue

\EndIf 

\EndWhile

\end{algorithmic}

\end{algorithm}


\section{Case Study: Verifying a traffic intersection with expected differential privacy} \label{sec:case}

In this section, we apply SMC with EDP (Algorithm~\ref{alg:2}) to a PTV Vissim traffic flow simulator of the intersection at West University Avenue and 13th Street at Gainesville, Florida, USA (Fig.~\ref{fig:vissim scene}). We analyze vehicles making driving decisions while crossing the intersection in the following order: turning right, driving straight, and turning left. Any parameters or calculations associated with the driving decisions are in the same order and the set of values are denoted with a text subscript ``veh.'' This case study focuses on the speed of cars crossing the intersection by observing the deviation percentage $e: [0,T] \rightarrow \real$ between the recorded speed $v: [0,T] \rightarrow \real$ and the speed limit $v_\text{lim} \in \{ 13, 50, 15 \}_\text{veh}$ as follows,
\begin{equation*}
    \left\{ e(t) = \frac{ v(t) - v_\text{lim} }{ v_\text{lim} } \right\}_i
\end{equation*}
over a simulation time horizon of $T \in \real$ seconds and where $i \in \{ 1, 2, 3 \}$ is the array index to denote the driving decisions turning right, driving straight, and turning left respectively. The vehicles' speed through the intersection is set at the following Gaussian distribution
\begin{equation}
    \left\{ v \sim \text{Gaussian}(v_\text{lim},\sigma^2) \right\}_i
    \label{eq: speed gaussian}
\end{equation}
where $\sigma \in \{ 3, 10, 5 \}_\text{veh}$. Finally, the traffic volume rate is set at 300 vehicles per hour per inbound road and the probability distribution of driving decisions follows $\{ 0.2, 0.6, 0.2 \}_\text{veh}$.



The requirement for $e$ is to enter a desired region $|e|<0.20$ within the time interval $[0,T]$ because vehicle speeds greater than the speed limit will endanger pedestrians while speeds less than the limit will impede traffic flow. For a given traffic flow $\mathcal{S}$, we want to check if this requirement holds with a probability greater than a desired threshold $p$.
Formally, in the STL syntax mentioned in Section~\ref{sec:smc}, we are interested in checking the following specification:
\begin{equation} \label{eq:traffic STL}
    \left\{ \mathbb{P}_\sigma \big( \sigma \models \Diamond_{[0, T]}(|e| < 0.20) \big) > p \right\}_i
\end{equation}
where $T=4$ minutes and $p \in \{ 0.50, 0.35, 0.34 \}_\text{veh}$. Evaluations are performed on a desktop with Intel Core i7-10700 CPU @ 2.90 GHz and 16 GB RAM. The code can be found at \url{https://github.com/SmartAutonomyLab/SMC-EDP}.

\textit{Results Analysis.}
Algorithm~\ref{alg:2} was used to analyze the traffic model with different combinations of the significance level $\alpha \in \{0.01, 0.05\}$, indifference parameter $\delta \in \{0.01, 0.03\}$, and privacy parameter $\varepsilon \in \{0.01, 0.05\}$ for vehicles turning right, driving straight, and turning left through the intersection. 

We estimated the satisfaction probability $p_\varphi \in \{ 0.64, 0.50, 0.49 \}_\text{veh}$ with a standard deviation of about $\{ 0.05, 0.05, 0.05 \}_\text{veh}$ using $10^4$ random samples. Thus, Assumption \ref{as:indifference} for implementing Algorithm~\ref{alg:2} holds since $p_\varphi - p \in \{ 0.14, 0.15, 0.15 \}_\text{veh}$, which is approximately $\{ 2.80, 3.00, 3.00 \}_\text{veh}$ times the standard deviation and is greater than both values of $\delta$ considered.
Additionally, we know from the estimated $p_\varphi$ that specification \eqref{eq:traffic STL} is true, so Algorithm~\ref{alg:2} should return $(H_{\text{null}})$ with probability of at least $1 - \alpha$.

On MATLAB, Algorithm~\ref{alg:2} ran $N = 10^4$ times for each of the considered combination of parameters for each driving decision. Then, we calculated the algorithm's accuracy (Acc.) for each combination, which is defined as follows:
\begin{equation} \label{eq:accuracy}
    \text{Acc.} := \frac{1}{N}\sum_{i=1}^{N} \mathbf{I}( o_i = H_{\text{null}}),
\end{equation}
where $\mathbf{I}$ is the indicator function and $o_i$ is the output of the $i^\mathit{th}$ run. In addition, the average sample termination time ($\tau_\mathscr{B}$), the average span of computation in minutes (Span), and the predicted hypothesis ($H_{\text{null}}$ or $H_{\text{alt}}$) were computed and presented in Table~\ref{tb: traffic results} for vehicles turning right, driving straight, and turning left through the intersection.

\begin{figure}[t]
    \centering
    \vspace{5pt}
    \includegraphics[width=0.9\linewidth]{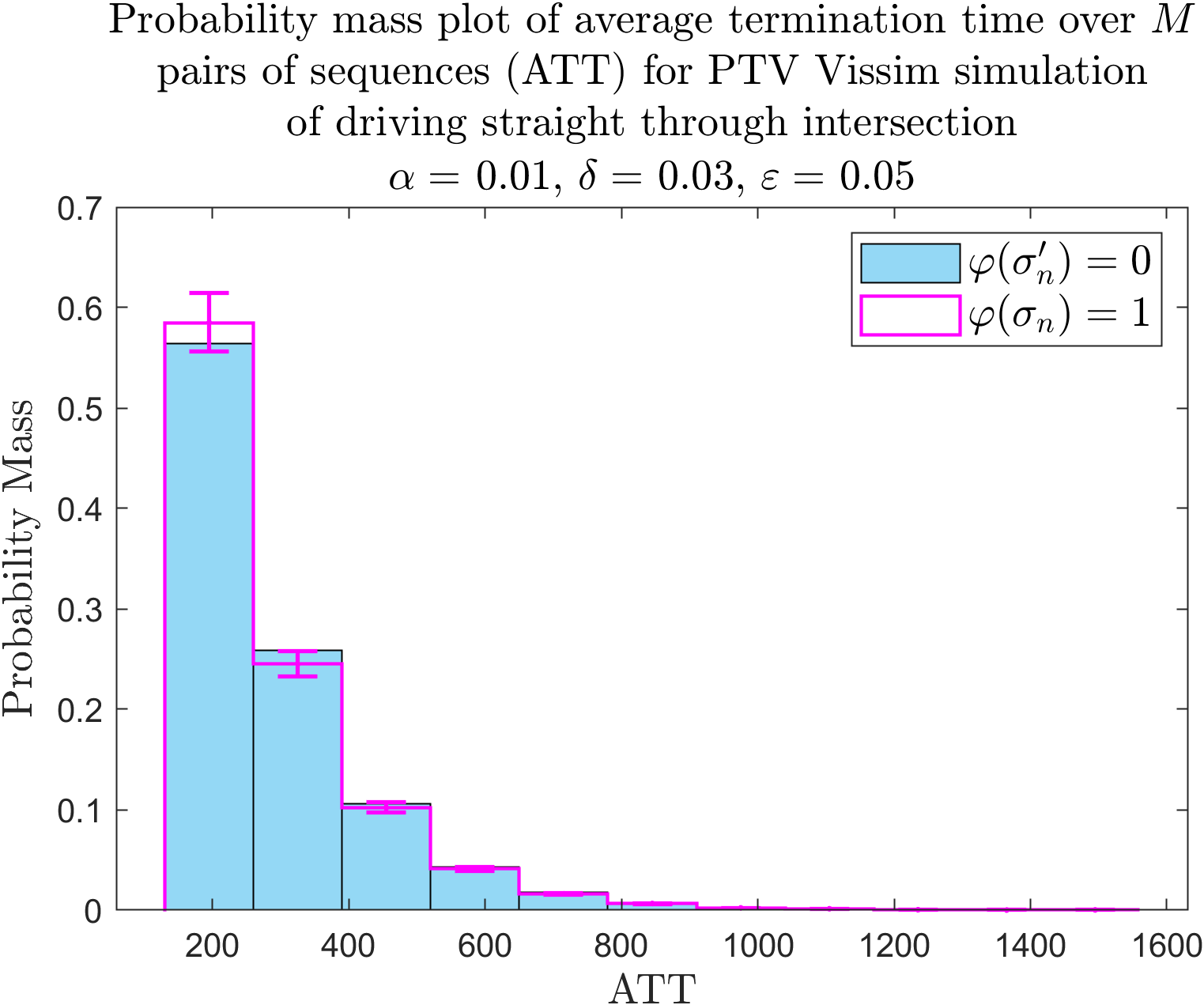}
    \caption{\footnotesize Empirical distributions of the average termination time over $M$ pairs of sequences (ATT) due to randomness from Algorithm~\ref{alg:2} for two input sequences of signals $\sigma_{1:\infty}$ and $\sigma_{1:\infty}'$. ATT is defined by \eqref{eq:ATT}, bins of ATT (with bin width $=130$) are on the $x$-axis, and the probability of an ATT falling within a bin is on the $y$-axis. The magenta border is the ATT distribution of $\sigma_{1:\infty}$ with satisfaction $\varphi(\sigma_n)=1$. The light blue region is the ATT distribution of $\sigma_{1:\infty}'$ with satisfaction $\varphi(\sigma_n')=0$. The magenta error bars are the tolerated change by factors of $e^{-\varepsilon}$ and $e^\varepsilon$ for differential privacy (i.e. it is the product of the probability for the ATT of $\sigma_{1:\infty}$ being in a specific bin with $e^{-\varepsilon}$ and $e^{\varepsilon}$, respectively). The closeness of the two distributions indicates differential privacy.}
    \label{fig:traffic probability mass plot}
    \vspace{-13pt}
\end{figure}

Afterwards, the differential privacy of Algorithm~\ref{alg:2} was analyzed for $M$ pairs of sequences of samples $\sigma_{1:\infty}$ and $\sigma_{1:\infty}'$ that differed in the $n^{\mathit{th}}$ entry.
Each entry in a sequence represents the satisfaction of $\varphi$ for the model sample, as defined in~\eqref{eq:varphi}. The average termination time over $M$ pairs of sequences (ATT) is as follows:

\begin{equation} \label{eq:ATT}
    \text{ATT} := \frac{1}{M}\sum_{j=1}^{M} \tau_{\mathscr{B}}^{(j)}(\cdot),
\end{equation}
where $M=500$ and $\tau_{\mathscr{B}}^{(j)}(\cdot)$ represents the termination time for the $j^\mathit{th}$ sample value of $\sigma_{1:\infty}$ and $\sigma_{1:\infty}'$. The ATT was calculated for each of $10^4$ samples of $L$, as defined in equation~\eqref{eq:randomized stopping condition}, and the resulting distributions can be seen for one of the parameter combinations in Fig.~\ref{fig:traffic probability mass plot}.

\begin{table}[t]
    \vspace{5pt}
    \caption{\small Results of statistical model checking with expected differential privacy for vehicles turning right, driving straight, and turning left through the traffic intersection. Various combinations of significance level $\alpha$, indifference parameter $\delta$, and privacy parameter $\varepsilon$ were analyzed. The algorithm's accuracy (Acc.) in predicting the null hypothesis ($H_{\text{null}}$) is defined by \eqref{eq:accuracy}. Additionally, the algorithm's average termination time ($\tau_\mathscr{B}$) and average span of computation in minutes (Span) are also presented here. The margins of error were computed using two standard deviations of the samples. Notice that increasing the value for any of the three parameters will subsequently reduce the average termination time and the average span of computation.}
    \label{tb: traffic results}
    \vspace{-4pt}
    \begin{subtable}{0.45\textwidth}
        \caption{\scriptsize Vehicles turning right}
        \vspace{-5pt}
        \label{tb: right}
        \centering
        \begin{tabular}[c]{ c c c c r r c }
            $1 - \alpha$ & $\delta$ & $\varepsilon$ & Acc. & $\tau_\mathscr{B}$ $(\times 10^3)$ & Span (min.) & $H_{\text{null}}$\\
            \hline
            $0.99$ & $0.01$ & $0.01$ & $1.00$ & $1.10 \pm 0.02$ & $73.51 \pm 1.09$ & T\\
            $0.99$ & $0.01$ & $0.05$ & $1.00$ & $0.54 \pm 0.00$ & $36.25 \pm 0.22$ & T\\
            $0.99$ & $0.03$ & $0.01$ & $1.00$ & $0.83 \pm 0.02$ & $55.50 \pm 1.09$ & T\\
            $0.99$ & $0.03$ & $0.05$ & $1.00$ & $0.28 \pm 0.00$ & $18.43 \pm 0.22$ & T\\
            $0.95$ & $0.01$ & $0.01$ & $1.00$ & $0.96 \pm 0.02$ & $64.00 \pm 1.10$ & T\\
            $0.95$ & $0.01$ & $0.05$ & $1.00$ & $0.40 \pm 0.00$ & $26.63 \pm 0.22$ & T\\
            $0.95$ & $0.03$ & $0.01$ & $1.00$ & $0.79 \pm 0.02$ & $52.60 \pm 1.08$ & T\\
            $0.95$ & $0.03$ & $0.05$ & $1.00$ & $0.23 \pm 0.00$ & $15.15 \pm 0.22$ & T
        \end{tabular}
        \vspace{5pt}
    \end{subtable}
    \begin{subtable}[c]{0.45\textwidth}
        \caption{\scriptsize Vehicles driving straight}
        \vspace{-5pt}
        \label{tb: straight}
        \centering
        \begin{tabular}{ c c c c r r c }
            $1 - \alpha$ & $\delta$ & $\varepsilon$ & Acc. & $\tau_\mathscr{B}$ $(\times 10^3)$ & Span (min.) & $H_{\text{null}}$\\
            \hline
            $0.99$ & $0.01$ & $0.01$ & $1.00$ & $1.02 \pm 0.02$ & $67.72 \pm 1.04$ & T\\
            $0.99$ & $0.01$ & $0.05$ & $1.00$ & $0.48 \pm 0.00$ & $32.27 \pm 0.21$ & T\\
            $0.99$ & $0.03$ & $0.01$ & $1.00$ & $0.78 \pm 0.02$ & $52.09 \pm 1.04$ & T\\
            $0.99$ & $0.03$ & $0.05$ & $1.00$ & $0.25 \pm 0.00$ & $16.59 \pm 0.21$ & T\\
            $0.95$ & $0.01$ & $0.01$ & $1.00$ & $0.89 \pm 0.02$ & $59.43 \pm 1.04$ & T\\
            $0.95$ & $0.01$ & $0.05$ & $1.00$ & $0.36 \pm 0.00$ & $23.85 \pm 0.21$ & T\\
            $0.95$ & $0.03$ & $0.01$ & $1.00$ & $0.74 \pm 0.02$ & $49.32 \pm 1.04$ & T\\
            $0.95$ & $0.03$ & $0.05$ & $1.00$ & $0.21 \pm 0.00$ & $13.79 \pm 0.21$ & T
        \end{tabular}
        \vspace{5pt}
    \end{subtable}
    \begin{subtable}[c]{0.45\textwidth}
        \caption{\scriptsize Vehicles turning left}
        \vspace{-5pt}
        \label{tb: left}
        \centering
        \begin{tabular}{ c c c c r r c }
            $1 - \alpha$ & $\delta$ & $\varepsilon$ & Acc. & $\tau_\mathscr{B}$ $(\times 10^3)$ & Span (min.) & $H_{\text{null}}$\\
            \hline
            $0.99$ & $0.01$ & $0.01$ & $1.00$ & $1.02 \pm 0.02$ & $68.20 \pm 1.05$ & T\\
            $0.99$ & $0.01$ & $0.05$ & $1.00$ & $0.49 \pm 0.00$ & $32.60 \pm 0.21$ & T\\
            $0.99$ & $0.03$ & $0.01$ & $1.00$ & $0.79 \pm 0.02$ & $52.54 \pm 1.04$ & T\\
            $0.99$ & $0.03$ & $0.05$ & $1.00$ & $0.25 \pm 0.00$ & $16.90 \pm 0.21$ & T\\
            $0.95$ & $0.01$ & $0.01$ & $1.00$ & $0.90 \pm 0.02$ & $60.14 \pm 1.05$ & T\\
            $0.95$ & $0.01$ & $0.05$ & $1.00$ & $0.36 \pm 0.00$ & $24.17 \pm 0.21$ & T\\
            $0.95$ & $0.03$ & $0.01$ & $1.00$ & $0.75 \pm 0.02$ & $50.28 \pm 1.07$ & T\\
            $0.95$ & $0.03$ & $0.05$ & $1.00$ & $0.21 \pm 0.00$ & $14.06 \pm 0.21$ & T
        \end{tabular}
        \vspace{5pt}
    \end{subtable}
    \vspace{-16pt}
\end{table}

\textit{Discussion.}
In Table \ref{tb: traffic results}, each parameter combination yielded an accuracy of $1.00$, which agrees with the confidence level $1 - \alpha$. The high accuracy results from choosing a small indifference parameter $\delta$, which increases the termination time and thus the number of samples. This can be seen in the likelihood ratio definition \eqref{eq:probability_ratio} where increasing $\delta$ (while satisfying Assumption~\ref{as:indifference}) will reduce the termination time. Furthermore, relaxing $1 - \alpha$ from $0.99$ to $0.95$ while holding the indifference parameter $\delta$ and privacy level $\varepsilon$ constant reduced the average sample termination time needed to output the null hypothesis.

The effects of increasing privacy level while holding $1 - \alpha$ and $\delta$ constant can be seen in both Table \ref{tb: traffic results} and Fig. \ref{fig:traffic probability mass plot}. Increasing $\varepsilon$ concentrated the distribution of $L$ in Algorithm~\ref{alg:2}, which led to less randomness in the sample termination time and thus, decreased the sample privacy.
Meaning, increasing $\varepsilon$ will decrease the average termination time, but will also make it easier to infer user data from the sample distribution and SMC output.

\section{Conclusion} \label{sec:conc}

This work used statistical model checking (SMC) with differential privacy to verify traffic models. Due to complexities and uncertainties associated with traffic systems, it is difficult to verify traffic model performance with traditional model-based approaches. SMC overcomes this obstacle by drawing samples from the system until a specification can be inferred with the desired confidence level. However, SMC may unintentionally leak sensitive traffic data when intruders observe the algorithm's output and termination time. Thus, we introduced expected differential privacy (EDP) and incorporated an exponential randomization mechanism into the SMC algorithm to achieve EDP. The modified algorithm was used in a PTV Vissim traffic model case study to demonstrate its accuracy in verifying specifications and its ability to keep data private. 

\bibliography{yu}

\bibliographystyle{IEEEtran}

\end{document}